\documentclass[sigconf]{acmart}
\usepackage{multirow}

\AtBeginDocument{%
  \providecommand\BibTeX{{%
    \normalfont B\kern-0.5em{\scshape i\kern-0.25em b}\kern-0.8em\TeX}}}

\copyrightyear{2021}
\acmYear{2021}
\setcopyright{rightsretained}
\acmConference[ACI'21]{Eight International Conference on Animal-Computer Interaction}{November 8--11, 2021}{Bloomington, IN, USA}
\acmBooktitle{Eight International Conference on Animal-Computer Interaction (ACI'21), November 8--11, 2021, Bloomington, IN, USA}
\acmDOI{10.1145/3493842.3493902}
\acmISBN{978-1-4503-8513-8/21/11}

\begin{document}

\title{Alexa, Play Fetch! A Review of Alexa Skills for Pets}

\author{Justin Edwards}
\affiliation{%
  \institution{University College Dublin}
  \department{School of Information and Communication Studies}
  \city{Dublin}
  \country{Ireland}
  }
\email{justin.edwards@ucdconnect.ie}

\author{Orla Cooney}
\affiliation{%
  \institution{University College Dublin}
  \department{School of Information and Communication Studies}
  \city{Dublin}
  \country{Ireland}}
\email{orla.cooney@ucdconnect.ie}

\author{Rachel Edwards}
\affiliation{%
  \institution{University College Dublin}
  \department{School of Veterinary Medicine}
  \city{Dublin}
  \country{Ireland}}
\email{rachel.edwards2@ucdconnect.ie}

\renewcommand{\shortauthors}{Edwards, et al.}

\begin{abstract}
Alexa Skills are used for a variety of daily routines and purposes, but little research has focused on a key part of many people's daily lives: their pets. We present a systematic review categorizing the purposes of 88 Alexa Skills aimed at pets and pet owners and introduce a veterinary perspective to assess their benefits and risks. We present 8 themes of the purposes for Skills aimed at pets and their owners: Calming, Animal Audience, Smart Device, Tracking, Training and Health, Translator, Entertainment/Trivia, and Other - Human Audience. Broadly, we find that these purposes mirror the purposes people have for using Alexa overall, and they largely are supported by veterinary evidence, though caution must be used when Skills relate to animal health. More collaboration between Conversational Agent researchers and animal scientists is called for to better understand the efficacy of using Alexa with pets.
\end{abstract}




\maketitle

\section{Introduction}

As Conversation Agents become increasingly ubiquitous, more and more people are adopting the systems into their daily routines, with many leading smartphone models now containing some version of conversational agent (CA) built in. CAs allow users to interact with and navigate the digital world easily in hand-free scenarios, often also being a source of entertainment for many. This is true of the home as well, with CA devices marketed for home use increasing in popularity, including Amazon’s Echo speaker range housing it’s Conversational Agent ‘Alexa’. At home, people use CAs for a range of tasks. \citet{ammari_music_2019} found that users largely utilize these devices to streamline their existing day-to-day routines, with popular uses including; the playing and controlling of music, hands-free internet search and the control of Internet of Things (IoT) devices embedded in the home; smart lights, thermostats, and smart security camera systems amongst others. In relation to IoT devices, \citet{mennicken_hacking_2012} note that ‘a convenient system is one that “fits, speeds up, or improves” family routines’. Previous research by \citet{porcheron_voice_2018} echoes this finding, suggesting that CAs become enmeshed within users’ home life. The systems allow users to complete tasks while at the same time engaging in regular home activities, such as eating dinner together as a family. They note that this apparent integration into the household is due in part to the continuous availability of VUIs through a simple wake word \cite{porcheron_voice_2018}. Research also suggests that conversational agents are more likely to be personified by users when the device is situated in a multi-member household or family unit \cite{purington_alexa_2017}. These studies, along with wider research, suggests that users are adopting conversational agents into the midst of their households. 

With CA devices seemingly reaching every corner of home life, It is perhaps unsurprising that a market has emerged for Alexa ‘Skills’ targeted towards pet owners. When we consider the fact that pets are often seen as family members themselves and are integral parts of many peoples’ home life, this appears to be a natural next step. Care of a pet is a major aspect of many peoples’ daily routine, whether it be feeding, walking, or grooming the pet among numerous other daily responsibilities, and these commitments have a major impact on the happiness of both pet and owner \cite{holland_acquiring_2019, bouma_expectations_2020}. Some work has called for a greater understanding of human and animal interactivity with agents \cite{moore_vocal_2016}, but this topic is underexplored, and we don’t currently know much about the interactions already taking place in people’s homes.

 With Alexa Skill functions ranging from the trivial, such as the Skill that will ‘marry’ your cats for you\footnote{amazon.com/Sarah-Dunlap-Cat-Wedding/dp/B07NMP17T8}, to the more practical, such as Skills promising trust-worthy pet health advice\footnote{ amazon.com/Vet24seven-Inc-MyPetDoc/dp/B07FP2N457
}, key questions remain around the scope of these Skills and how beneficial they could really be. This paper aims to explore these questions by offering an initial overview of the types of Alexa Skills people use with their pets and the their supposed purposes. We also introduce a veterinary perspective to discuss the potential benefits, or risks, of using a CA in these ways with animals, highlighting some of the Alexa Skills currently available for use with pets. 

\section{Methods}
We searched the Alexa Skills category on Amazon.com for relevant English-language Skills in January 2021 using the following search terms: “dog(s)”, “cat(s), and “pet(s)”. Each term was searched individually in its singular and plural form. Our initial search yielded 1,851 Skills. Screening for duplicates yielded 589 Skills. The lead author then screened the Skills based on the following criteria: 1) Only Skills with at least 5 user reviews were included to ensure all Skills were used by Alexa users. 2) Only Skills which are explicitly aimed at pets or  pet owners were included (examples of other Skills that were excluded are Skills that give a user facts about animals, Skills that play animal noises, and Skills that coincidentally have “cat” or “dog” in a product name). After this screening, we arrived at our final 88 Skills for analysis.

The lead author and a co-author conducted inductive Thematic Analysis \cite{braun2006using} to categorize the purpose for each of the 88 Skills. Initial themes were generated by the lead author and then independently used for categorization by the co-author. After initial coding, there was 87\% agreement between the two authors, and inconsistencies were resolved through discussion. The themes of purposes for these Skills are summarized below and a complete table of Skills, descriptions, and themes is included in supplementary materials.

\section{Themes}
\subsection{Calming} 
This theme included Skills which are intended to be heard by a pet for the purpose of calming or relaxing the pet. This included Skills like Dog Sleep Music\footnote{amazon.com/Simmba-Dog-Sleep-Music/dp/B0859K698K}, Calm My Dog\footnote{amazon.com/Stephen-Brown-Calm-My-Dog/dp/B07G4B6WL4}, and Comfort My Dog\footnote{amazon.com/Voice-Games-Relax-My-Dog/dp/B07JYYHV5L}. These Skills use music or ambient noise to calm the listening pet. Some veterinary research has shown that playing classical music in shelter environments can help to calm dogs \cite{kogan_behavioral_2012}, so these Skills aimed at using music or ambient sound to sooth pets may likewise have the intended beneficial effect.  

\subsection{Animal Audience} 
This theme included Skills which are intended to be heard by a pet for them to react (commands, calls, toys, other noises) but not in the calming category below, including Skills like Make My Dog Howl\footnote{amazon.com/Iguana-ASD-Make-Dog-Howl/dp/B07XC4JMMS} and Calling Your Dog\footnote{amazon.com/Jobless-Calling-Your-Dog/dp/B08BZ8626H}. These Skills intend to entertain or stimulate pets, often by simulating the sounds of other animals or of people. This may be another area in which the Skills have their indented effect, as some research has indicated that dogs are stimulated by watching and hearing other dogs on television \cite{hirskyj-douglas_dog_2017}. 

\subsection{Smart Device}
Some Skills are intended as a mode of interaction with smart devices, like the Dog Whisperer\footnote{amazon.com/Mario-Harper-Dog-Whisperer/dp/B01MXX19JL} Skill which connects with the FitBark device and the PetSafe® Smart Feed Skill\footnote{amazon.com/PetSafe-PetSafe\%C2\%AE-Smart-Feed/dp/B07RRSW2SV} which controls the PetSafe automatic feeder. The devices these Skills are linked with are primarily aimed at automatically feeding pets and monitoring their activity, which has been shown to produce positive health outcomes in a recent study involving multi-cat households \cite{lambrecht_can_2019}. That said, the efficacy of a smart devices is dependent on the capabilities of the device, the behaviour of the pet, and the integration of the device into the specific context of the household \cite{lambrecht_can_2019}, so the use of such a Skill and a smart device is no guarantee of its utility to pet or owner.

\subsection{Tracking}
A related theme, tracking Skills, likewise help pet owners monitor their pets feeding and exercise, but via manual tracking rather than using a smart device. These Skills, like Cat Feed Tracker\footnote{amazon.com/Kuske-Cat-Feed-Tracker/dp/B01MT3WAUS} and Pet Tracker\footnote{amazon.com/Doug-Johnson-Pet-Tracker/dp/B077M2X3VS} allow users to track their pet care and check if other users have logged any pet activities. The multi-user aim fits with prior work on smart speakers like Amazon Echo which has indicated that a single device is frequently used by multiple members of a home \cite{porcheron_voice_2018}. Likewise, veterinary research indicates consistent routines for feeding and exercise are beneficial to pet health \cite{vitger_integration_2016}. To the extent these routines can be maintained through Alexa Skills, this usage pattern may mirror the integration of Alexa into family routines that has been observed in other contexts. \cite{porcheron_voice_2018}.

\subsection{Training and Health}
Other Skills, like Doctor Pupper\footnote{amazon.com/melochi-Doctor-Pupper/dp/B07K7CSRTP}, My Pet Doc\footnote{amazon.com/Vet24seven-Inc-MyPetDoc/dp/B07FP2N457}, and Al’s Dog Training Tips\footnote{amazon.com/Longoriahaus-Dog-Training-Als-Tips/dp/B07H2WMDTV} aim to help pet owners with tips on training and with simple medical advice such as advice on whether certain foods are toxic to pets. These Skills have good intentions, particularly as most pet poisonings result from owners sharing foods they don’t know are harmful \cite{gugler_hidden_2013}, but research on conversational technology like Alexa giving medical advice has illustrated the severe harm that can be caused by technical errors and limited safeguards against bad advice \cite{bickmore_patient_2018}. 

\subsection{Translator}
Many Skills, like Dog Translator\footnote{amazon.com/Steven-Foyston-Dog-Translator/dp/B088PK52LL} and Cat Translator\footnote{amazon.com/GeeNelly-Cat-Translator/dp/B079FPVGLC} purport to listen to a pet and explain to owners what their pets are saying. While there is some evidence that sounds like dog barks can convey different emotional states \cite{pongracz_acoustic_2006}, it is not clear that the Skills available presently can actually detect these acoustic differences. These Skills should not be considered serious utilities to pet owners.

\subsection{Entertainment/Trivia}
Several Skills like Name My Dog\footnote{amazon.com/BethSherm-Name-my-dog/dp/B0741F6WXR} or Cat Wedding\footnote{amazon.com/Sarah-Dunlap-Cat-Wedding/dp/B07NMP17T8} are intended purely as entertainment for pet owners, with activities that involve their pet or directly relate to owners’ relationships with their pets. These Skills don't purport to have any benefit to pets or their owners beyond entertainment of the owner.

\subsection{Other - Human Audience}
Finally, some Skills did not fit into the above categories but were aimed at human users and their relationships with animals and had practical purposes. Skills in this theme included PawBoost Lost and Found Pets\footnote{amazon.com/PawBoost-Lost-and-Found-Pets/dp/B01MQW0BC0} which lets users communicate about lost pets and Pet Finder\footnote{amazon.com/Monika-Wiest-Pet-Finder/dp/B01N7M3DKY} which guides a user through questions about their lifestyle to suggest suitable dog breeds for adoption. 

\section{Discussion}
People use voice assistants like Alexa for a diverse array of purposes, often pertaining to tasks like playing music, searching the web, and interacting with smart devices \cite{ammari_music_2019}. In our review of Alexa Skills for pets and pet owners, we found a very similar pattern of purposes for using Alexa. Broadly, the themes we present, Calming, Animal Audience, Smart Device, Tracking, Training and Health, Translator, Entertainment/Trivia, and Other - Human Audience represent a variety of strategies for caring for pets, strengthening our bonds with them, and keeping them entertained. People use Alexa for their pets largely in the same ways that they use Alexa in the home setting in general; as an extension of existing routines (feeding, health tracking, training), with a number of these Skills offering straightforward support for pet owners’ creation and maintenance of routines. Some Skills likely do not or cannot technically achieve their desired outcome (i.e. Skills aimed at translating pet sounds) or offer minimal utility beyond entertaining pet owners. However, for many of the themes themes of Skill purposes we reviewed, there is veterinary evidence supporting benefits of the Skills' intention, and reason to believe those purposes can be acheived through an Alexa Skill. 

Pet owners should take caution that following medical advice that comes from voice assistants like Alexa is risky \cite{bickmore_patient_2018} and that individual pets require different approaches to care and training \cite{turcsan_trainability_2011}. The integration of technology into pet-care routines is not one-size-fits-all. Smart devices and technological aids to pet care must fit well into the structure of a pet's home, creating an ongoing challenge for both animal-computer interface designers and pet owners \cite{lambrecht_can_2019}.  Still, most pet-focused Alexa Skills treat Alexa as an enhancement to a human-animal relationship, rather than a replacement for it. Insofar as pet owners see these Skills this way, as toys and tools at their hand as pet owners, but not as a replacement for training or veterinary expertise, the purposes for these Skills seem justifiable. Many questions remain in the wider field of CAs for animal-computer interaction, with further research into their efficacy posing an interesting point of collaboration between human-computer interaction and animal science researchers. We aim to offer a first step in investigating Alexa Skills for pets, surveying the Skills that currently are in use. Future work should deepen our understanding of this topic by seeking to understand the outcomes for the pets and the experiences of pet owners who use such Skills.

The video poster accompanying this extended abstract can be found at https://youtu.be/H30qRRbrY68
\begin{acks}
%Acknowledgements withheld for review
This research was conducted with the financial support of the ADAPT Science Foundation Ireland (SFI) Research Centre at University College Dublin and the SFI Centre for Research Training in Digitally Enhanced Reality (D-REAL). The ADAPT SFI Centre for Digital Content Technology is funded by Science Foundation Ireland through the SFI Research Centres Programme and is co-funded under the European Regional Development Fund (ERDF) through Grant No. 13/RC/2106\textunderscore{}P2 and D-REAL funding is provided under SFI under Grant No. 18/CRT/6224.
\end{acks}

\bibliographystyle{ACM-Reference-Format}
\bibliography{pets}


\end{document}